\documentclass[a4paper,10pt,oneside]{article}
\usepackage{epsfig}
\usepackage{hangcaption}
\usepackage{ngerman}
\usepackage{graphicx}
\usepackage[bf]{caption}
\usepackage[dvips]{hyperref}
\usepackage{fancyhdr}
\newcommand{\qqbar}{\antibar{q}}
\newcommand{\ttbar}{\antibar{t}}
\newcommand{\antibar}[1]{\mathrm{#1\overline{#1}}}
\newcommand{\RTPC}{\mathrm{R}_{\mathrm{TPC}}}
\newcommand{\LTPC}{\mathrm{L}_{\mathrm{TPC}}}
\newcommand{\SIGMACENT}{\sigma_{\mathrm{c}}}
\newcommand{\RMS}{\mathrm{RMS}_{90}}
\newcommand{\DESY}{\mbox{DESY}}
\newcommand{\ILC}{\mbox{ILC}}
\newcommand{\LHC}{\mbox{LHC}}
\newcommand{\LDC}{\mbox{LDC}}
\newcommand{\LCIO}{\mbox{\tt LCIO}}
\newcommand{\GEAR}{\mbox{\tt Gear}}
\newcommand{\GEANT}{\mbox{\tt Geant4}}
\newcommand{\MOKKA}{\mbox{\tt Mokka}}
\newcommand{\MARLIN}{\mbox{\tt Marlin}}
\newcommand{\MARLINRECO}{\mbox{\tt MarlinReco}}
\newcommand{\MARLINUTIL}{\mbox{\tt MarlinUtil}}
\newcommand{\CEDVIEWER}{\mbox{\tt CEDViewer}}
\newcommand{\PANDORAPFA}{\mbox{\tt PandoraPFA}}
\newcommand{\MAGIC}{\mbox{\tt Magic}}
\newcommand{\ROOT}{\mbox{\tt ROOT}}
\newcommand{\AIDA}{\mbox{\tt AIDA}}
\newcommand{\RAIDA}{\mbox{\tt RAIDA}}
\newcommand{\TASSO}{\mbox{Tasso}}
\newcommand{\JETSET}{\mbox{JETSET}}
\newcommand{\GRID}{\mbox{GRID}}
\newcommand{\FORTRAN}{\mbox{Fortran}}
\newcommand{\BRAHMS}{\mbox{\tt Brahms}}
\newcommand{\NIMA}[3] {Nucl.\ Instr.\ and Meth.\ \textbf{A#1} (#2) #3}
\def\etal{\mbox{{\it et al.}}}

\hypersetup{pdfhighlight=/O,pdfpagemode=None,pdfstartview=XYZ,pdftitle={Event Reconstruction with {MarlinReco} at the ILC},pdfauthor={O.~Wendt},pdfsubject={MarlinReco},pdfkeywords={linear collider, simulation, software tools, event reconstruction, particle flow}}

\title{\textbf{Event Reconstruction with {MarlinReco} at the ILC\footnote{to appear in Proceedings of LCWS06, Bangalore, India, March 2006}}}
\author{O.~Wendt$^{1,2}$, F.~Gaede$^{1}$, T.~Kr\"amer$^{1}$}

\date{}

\begin{document}
\maketitle

\vspace{-5ex}

\begin{table}[h]
\centering	
\begin{tabular}{r@{ }l}
$^{1}$ & DESY, Notkestrasse~85, 22607~Hamburg, Germany\\
%       & \\
$^{2}$ & Universit\"at~Hamburg, Institut f\"ur Experimentalphysik,\\ 
       &  Luruper~Chaussee~147, 22761~Hamburg, Germany \\
%       & \\
       &
\end{tabular}
\end{table}

\fancyhead{} % clear all fields
\fancyhead[R]{}
\fancyhead[L]{}

\begin{abstract}\noindent
After an overview of the modular analysis and reconstruction framework
\MARLIN\ an introduction on the functionality of the \MARLIN-based
reconstruction package \MARLINRECO\ is given. This package includes a
full set of modules for event reconstruction based on the Particle
Flow approach. The status of the software is reviewed and recent
results using this software package for event reconstruction are
presented.
\phantom{Keywords:}~particle~flow
\newline
\newline
\end{abstract}

\section{Introduction}
The International Linear Collider (\ILC) will be the next machine
beyond the \LHC. It allows to explore the physics at the 500~GeV to
1~TeV energy scale with high precision. Sophisticated simulation and
reconstruction software supports the ongoing development and
optimisation of a detector for the
\ILC. Fig.~\ref{fig:LDC_and_Marlin}~(a) shows a schematic overview of
the core software chain used for the studies of the Large Detector
Concept (\LDC), one of the four current detector proposals for the
\ILC~\cite{ldcwww}. This chain consists of two major parts:
\begin{enumerate}
\item{} The \GEANT-based simulation of the detector response, \MOKKA~\cite{geant4,ilcsoftportalwww}.
\item{} The digitisation of simulated data, event reconstruction as well as analysis provided by different modules of the \MARLIN~framework~\cite{ilcsoftportalwww,Gaede:2006pj}.
\end{enumerate}
The event data is shipped through the software chain using
\LCIO~\cite{Gaede:2003ip} between different programs. Geometry
related data needed by the reconstruction is provided by the full
detector simulation and can be accessed via
\GEAR~\cite{ilcsoftportalwww}.

\section{{Marlin} and {MarlinReco}}
\label{sec:Marlin_MarlinReco}
\begin{figure}[ht]
  \epsfxsize=11cm \centerline{\epsfbox{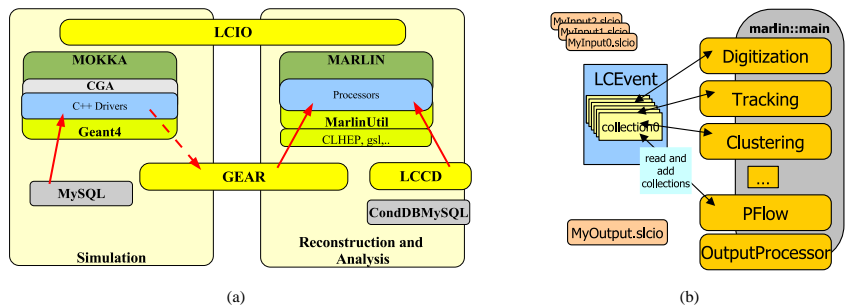}}
  \caption[\LDC\ simulation and reconstruction framework~(a),
  structure of Marlin~(b)]{\LDC\ simulation and reconstruction
  framework~(a), structure of Marlin~(b)}
  \label{fig:LDC_and_Marlin}
\end{figure}

\MARLIN\ is a modular C++ application framework based on \LCIO\ for
the analysis and reconstruction of \ILC\ data. \MARLIN\ provides the
main program with the event loop and a mechanism to call modules, so
called processors, to carry out specific tasks. These tasks can be as
simple as filling histograms or as complex as a full event
reconstruction. As an example, the full chain of an event
reconstruction with processors for digitisation, tracking etc. is
shown in Fig.~\ref{fig:LDC_and_Marlin}~(b). The framework is reading
data event by event creating an {\it LCEvent} which is used to hand
the data from processor to processor during a \MARLIN\ run. An {\it
LCEvent} consist of a set of collections holding specific objects like
hits, tracks, cluster etc. Processors only have the permission to read
and add information to ensure the consistency of the data. Program
steering is done via an XML file allowing to hand over processor
parameters, specifying the order of the processors or exchanging
processors without recompilation.
The package \MARLINRECO\ is a specific set of processors for a complete
event reconstruction system, based on the Particle Flow
concept. Version 00-02 contains the following processors:

\begin{description}
\item[Tracker Hit Digitisation:] For the Vertex system there are two
different digitisers available. A simple digitiser translates
simulated tracker hits into tracker hits, without modifications. A
more sophisticated digitiser takes the deposition and transfer of
charge in silicon as well as the geometry into
account~\cite{vertexDet}. In the Time Projection Chamber (TPC) a
Gaussian smearing of the simulated hit positions in $r$-$\phi$ and $z$
is done to account for the intrinsic chamber resolution. Their
parameters are obtained from \GEAR.
\item[Calorimeter Hit Digitisation:] There are two different
digitisers for the electromagnetic and hadronic calorimeter (ECAL and
HCAL). The first provides calibration, low energy hit rejection and
various sampling fractions for the different regions of the
calorimeter. The second has the capability to merge neighboring cells
into larger cells. This feature allows the variation of the cell size
in a simple way. Both digitisers are able to treat hits in analogue and
digital calorimeters.
\item[Tracking:] There are two tracking processors. The first is
based on algorithms taken from LEP providing full tracking in the TPC
with energy loss and multiple scattering~\cite{Behnke:2001gb}. The
hits of the inner silicon detectors can be included in the track fit,
using the tracks reconstructed in the TPC as seeds. The second
processor provides a stand-alone pattern recognition procedure for the
vertex detector~\cite{vertexDet}.
\item[Clustering:] One of the central parts in the Particle Flow
approach is a sophisticated procedure to assign calorimeter hits to
the proper reconstructed particle and to minimise the ``confusion''
between adjacent particles. This so-called clustering is done by the
``Trackwise Clustering''~\cite{trackwiseclustering} algorithm. It only
relies on spatial information of the calorimeter hits with minimal
dependence on the detector geometry. It is applicable to digital and
analogue calorimeters as well as to different detector designs.
\item[Particle Flow:] \MARLINRECO{}'s Particle Flow processor ``Wolf''
extrapolates the tracks into the calorimeter and matches them to
clusters by a proximity cut taking into account the detector
geometry. In addition, a simple particle identification is done by
calculating the fraction of energy in the ECAL and the HCAL. After
that, a collection of reconstructed particles is created where the
four momenta of charged particles are determined by the track
parameters. The four momenta of neutral particles are calculated from
the clusters only.
\item[Track and Cluster Cheater:] Processors allowing the assignment 
of hits to tracks and clusters, using Monte Carlo information only,
are provided. To obtain the track parameters either a simple helix
hypothesis is fitted to the tracker hits or the information is taken
from the Monte Carlo directly.

\item[Analysis:] There are processors to calculate the thrust axis
and value (\TASSO\ and \JETSET\
algorithm~\cite{Brandt:1964sa,Sjostrand:1993yb}) as well as the
sphericity and aplanarity of an event. In addition, a multi-algorithm
jet finding processor is available~\cite{satorujetfinder}.
\item[Calibration:] This processor calculates the calibration
constants for the ca\-lo\-ri\-me\-ter by the method proposed
in~\cite{abscalibration}. It is based on the energy conservation law
giving an upper limit for the energy sum in all the cells of the
calorimeter.
\end{description}

\MARLINRECO\ is based on the package \MARLINUTIL\ combining
utility and helper classes and by the client-server based event
display system \CEDVIEWER. \RAIDA, a \ROOT\ implementation of the
\AIDA\ interface, is available~\cite{Root,Aida}. Due to the modular
structure and the well defined data structures, alternative algorithms
(\MAGIC, \PANDORAPFA~\cite{Ainsley:2004iu,pandorapfa}) can easily be
included in \MARLINRECO. All software packages as well as more
detailed documentation can be accessed via~\cite{ilcsoftportalwww}.

\section{Event Reconstruction}
\label{sec:EventReconstruction}
Here the \MARLINRECO-based event reconstruction is tested and the
dependence of the performance of the Particle Flow on basic geometric
properties of the detector is studied. For this purpose the full detector
simulation using \MOKKA\ v05.04 and event reconstruction with
\MARLINRECO\ is done with four classes ($\gamma / \mathrm{Z}^{0}
\rightarrow \qqbar$, $\mathrm{WW}$ and $\mathrm{Zh} \rightarrow
\textrm{4~jets}$ as well as $\ttbar
\rightarrow \textrm{6~jets}$) of events at four different center-of-mass
energies (91.2, 360, 500 and 1000~GeV). Four different layouts of the
\LDC\ and two values of the magnetic field have been chosen to optimise the
detector. For the variation of the detector geometry two detector
models, LDC00Sc and LDC01Sc provided by \MOKKA, with different
sampling structures in the ECAL are chosen. For each model two sizes
of the TPC, determined by their outer radius ($\RTPC$) and length from
the nominal IP to the end plane of the TPC ($\LTPC$), are
constructed~((A) and (B) in Tab.~\ref{tab:DetectorGeometry}). This
results, together with the two values of the magnetic field (3,~4~T)
in eight detector layouts. Tab.~\ref{tab:DetectorGeometry} summarises
the available geometries.
\begin{table}[h]

\hskip4pc\vbox{\columnwidth=26pc
\begin{tabular}{l|l|l|l|l}
model & \multicolumn{2}{l|}{\bf LDC01Sc} & \multicolumn{2}{l}{\bf LDC00Sc} \\
variation & \multicolumn{1}{c}{(A)} & \multicolumn{1}{c|}{(B)} & \multicolumn{1}{c}{(A)} & \multicolumn{1}{c}{(B)} \\ \hline
$\RTPC$ (mm) & 1380 & 1580 & 1690 & 1890 \\
$\LTPC$ (mm) & 2000 & 2200 & 2730 & 2930
\end{tabular}
}
\caption[Layouts of the \LDC\ simulated with \MOKKA\ v05.04]{Layouts of the \LDC\ simulated with \MOKKA\ v05.04. The four detector geometries are available with a magnetic field of 3 and 4~T.}
\label{tab:DetectorGeometry}
\end{table}
Computing and data storage for simulation and reconstruction has been
done using \GRID\ resources. Meta information about the simulated data
as well as the logical names to access the files are available through
a database~\cite{ilcsoftportalwww}. For this study Monte Carlo
information has been used to perform the pattern recognition in the
tracking system. A helical fit is applied afterwards to obtain the
track parameters.

\begin{figure}[h]
  \epsfxsize=12cm \centerline{\epsfbox{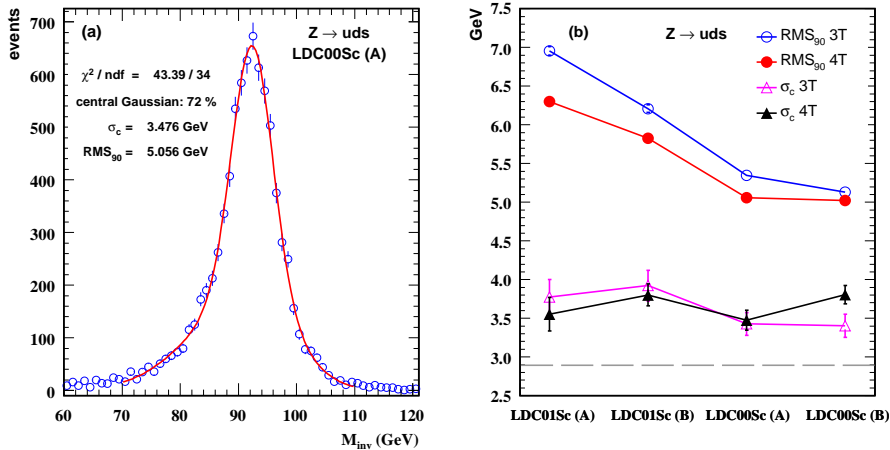}}
  \caption[Reconstructed invariant mass of $\mathrm{Z}^{0}\rightarrow{\mathrm{uds}}$ at $\sqrt{s} = \mathrm{m}_{\mathrm{Z}}$]{Reconstructed invariant mass of $\mathrm{Z}^{0}\rightarrow{\mathrm{uds}}$ at $\sqrt{s} = \mathrm{m}_{\mathrm{Z}}$ fitted with a sum of two Gaussians (a), performance of \MARLINRECO\ ($\RMS$ and $\SIGMACENT$) for different detector geometries (see Tab.~\ref{tab:DetectorGeometry}) and magnetic fields (b).}
  \label{fig:Z0Reco}
\end{figure}

In Fig.~\ref{fig:Z0Reco}~(a) the distribution of the invariant mass of
$\mathrm{Z}^{0}$ measured with $\mathrm{Z}^{0} \rightarrow
\mathrm{uds}$ events at $\sqrt{s} = \mathrm{m}_{\mathrm{Z}}$ for the
detector LDC00Sc~(A) with a magnetic field of 4~T is shown. Due to the
difference of the tails compared to a Gaussian distribution the
root-mean-square (RMS) is not an appropriate measure of the width of
the peak and therefore of the performance of the
reconstruction. Hence, (1)~the RMS is calculated with the bins around
the maximum bin containing 90\% of the
events~($\RMS$)~\cite{pandorapfa}, and (2) the sum of two Gaussian
functions, one for the central part and one for tails, is fitted to
the mass distribution. The width of the central
Gaussian~($\SIGMACENT$) is the measure of the width of the peak~(see
Fig.~\ref{fig:Z0Reco}~(a)). The results of both methods are quoted to
show the performance of the reconstruction. They also act as an
indicator in the process of detector optimisation. The results for the
geometry listed in Tab.~\ref{tab:DetectorGeometry} are shown in
Fig.~\ref{fig:Z0Reco}~(b).
To study the performance at higher energies, a simple analysis of
$\ttbar \rightarrow
\textrm{6~jets}$ at $\sqrt{s} = 500\ \mathrm{GeV}$ has been
performed by calculating $\Delta
\mathrm{E}_{\mathrm{reco}}:=\sum_{i}\mathrm{E}_{\mathrm{reco}}^{i}-\sum_{i}\mathrm{E}_{\mathrm{avail}}^{i}$
for each event. In the first part of $\Delta
\mathrm{E}_{\mathrm{reco}}$ the energies of all reconstructed
particles are added up, while in the second part the energy sum of all
Monte Carlo particles which pass the acceptance cut $\theta > 0.1$ and
which are not neutrinos is calculated. This results in
$\Delta~\mathrm{E}_{\mathrm{reco}}=25.2$~GeV which is about a factor
of two larger than the pure calorimeter resolution given by $\Delta
\mathrm{E}_{\mathrm{calo}}:=\sum_{i}\mathrm{E}_{\mathrm{calo}}^{i}-\sum_{i}\mathrm{E}_{\mathrm{avail}}^{i}=12.6$~GeV,
where the first part of $\Delta \mathrm{E}_{\mathrm{calo}}$ adds up
the energy of all calorimeter cells~\cite{abscalibration}. One reason
for this decrease of performance compared to $\mathrm{Z}^{0}
\rightarrow \mathrm{uds}$ at $\sqrt{s} = \mathrm{m}_{\mathrm{Z}}$ is
the misassignment of hits due to overlaps of showers in the
calorimeter. The \FORTRAN -based simulation and reconstruction package
\BRAHMS\ has shown that it is possible to reach energy resolutions of about 9~GeV
for $\ttbar$-events at $\sqrt{s} = 500\ \mathrm{GeV}$ following the
Particle Flow concept~\cite{ilcsoftportalwww,Chekanov:2003cp}.

\section{Conclusions}
For the determination of $\SIGMACENT$ for $\mathrm{Z}^{0} \rightarrow
\mathrm{uds}$ at $\sqrt{s} = \mathrm{m}_{\mathrm{Z}}$ \MARLINRECO\ comes close to the
performance goal of the jet energy resolution at the ILC ($\sigma_E/E
= 0.30/\sqrt{E}$ corresponding to $\sigma_E = 2.9\ \mathrm{GeV}$ at
$\sqrt{s} = m_{\mathrm{Z}}$) but no significant dependence on the
detector geometry is observed. The results of the $\RMS$-method are
considerably larger but are showing a clear dependence on the detector
geometry. In addition, this dependence follows the expectation,
i.e. the resolution increases with a larger detector and a larger
magnetic field~(see Fig.~\ref{fig:Z0Reco}~(b)). The analysis of
$\ttbar \rightarrow
\textrm{6~jets}$ at $\sqrt{s} = 500\ \mathrm{GeV}$ shows that
improvements in \MARLINRECO\ are necessary, especially for high
center-of-mass energies. Nevertheless, \MARLINRECO\ provides the full
chain of event reconstruction following the Particle Flow concept.
\\ \\{\bf  Acknowledgments}\\
We like to thank all members of the \DESY-FLC software group:
H.~Albrecht, S.~Aplin, T.~Behnke, P.~Krstonosic, D.~Martsch,
V.~Morgunov, J.~Samson, A.~Vogel

\end{document}